\documentclass{article}
\usepackage{hyperref}
\usepackage{epsfig,amssymb,amsmath,graphicx,lineno}
\def\gsim{\mathrel {\vcenter {\baselineskip 0pt \kern 0pt
    \hbox{$>$} \kern 0pt \hbox{$\sim$} }}}

\author{Esteban Roulet, for the Pierre Auger Collaboration\\
\small{\it CONICET, Centro At\'omico Bariloche, Av. Bustillo 9500 (8400) Argentina}\\
\small{\it Observatorio Pierre Auger, Av. San Mart\'in Norte, 304, 5613 , Malarg\"ue, Argentina}\\
\small Full author list: {\rm http://www.auger.org/archive/authors\_2014\_11.html} }

\title{Recent results from the Pierre Auger Observatory}
 
\date{}

\begin{document}

\maketitle

\begin{abstract}
 The results obtained after the first decade of operation of the Pierre Auger Observatory are reviewed.
\end{abstract}


\section{Introduction}

The Pierre Auger Observatory, located in Malarg\"ue, Argentina, is the largest existing air shower detector and has been taking data for more than 10 years, obtaining several fundamental results about the ultra-high energy cosmic rays (UHECRs). 

The main part of the Observatory is the surface detector, SD, consisting of an array of $\sim 1600$ water-Cherenkov detectors (WCD) separated by 1.5~km and covering an area of 3000~km$^2$. This array is fully efficient for air showers with energies $E>3$~EeV (where 1~${\rm EeV}  \equiv 10^{18}$~eV) and has a duty cycle close to 100\%.  
This array is overlooked by the fluorescence detector, FD, consisting of 27 telescopes located in 4 sites lying  on the periphery of the SD, which measure the fluorescence light emitted  by atmospheric nitrogen molecules that get excited by the air shower particles. It has a duty cycle of about 13\% since clear, moonless nights are required for these observations. The fluorescence technique is important because it allows one to obtain an almost calorimetric measurement of the shower energy and also allows one to determine the longitudinal development of the shower in the atmosphere, making it possible to infer the depth of maximum shower size, $X_{max}$, which provides relevant information on the nature of the primary CR particle. For instance, photons are on average more penetrating than hadrons, and among these last  the protons are on average more penetrating than the heavier nuclei. Moreover, since to a good approximation heavier nuclei of energy $E$ and mass number $A$ can be  considered as a collection of $A$ nucleons of energies $E/A$, the air showers they produce can be considered as averages of $A$ sub-showers. Hence, the depths of shower maximum they produce have considerably less fluctuations, $\sigma(X_{max})$, than the fluctuations between different proton induced showers.
The combined SD and FD detectors allow for hybrid measurements  of air showers with the two different techniques. This is used in particular to calibrate the energy estimator of SD, which is the signal at 1~km from the shower core obtained by fitting the lateral distribution of the measured signals in the WCDs. Hybrid measurements also allow for several cross-checks between the SD and FD  reconstructions.

Besides these detectors, about 1\% of the area is instrumented with a denser array of WCDs, the Infill, having a 750~m spacing between detectors. 
This sub-array is fully efficient down to $E\simeq 3\times 10^{17}$~eV and is used to explore this lower energy regime. In addition, the three HEAT telescopes look at higher elevations above the area of the Infill, being then sensitive also to lower energy air showers.

The Auger Collaboration consists of more than 400 scientists from 18 countries, and being that this talk is given at the Latin-American Symposium of High Energy Physics, SILAFAE, it is relevant to emphasize the significant Latin-American participation in it. Besides Argentina, which is  the host country, Brazil and M\'exico participate, and very recently also Colombia has joined as an associated country.

Some of the main issues that the Observatory aims to address are:
\begin{itemize}
\item To discover the origin of the UHECRs particles so as to better understand their sources and the acceleration mechanisms that produce them.

\item To understand the reasons for the changes in the spectrum, since these can provide information on the eventual transition from galactic to extragalactic sources, can be affected by the diffusive propagation of CRs across magnetic fields, and may be also shaped by the CR interactions with the photon backgrounds traversed during propagation from the sources to the Earth, etc.

\item To measure the CR composition to understand the reasons behind the changes observed.

\item To determine the anisotropies both on large scales (dipole and quadrupole), that could result from diffusion processes at lower energies or when many anisotropically distributed sources contribute to the fluxes,   and also on smaller angular scales that  could result at the highest energies for which the propagation from the nearby sources could become quasi-rectilinear.

\item To know whether neutrinos or photons are produced in association with the highest energy cosmic rays as they interact at the sources or during their propagation.

\item To learn about the hadronic interactions at energies well beyond those studied at accelerators.
\end{itemize}

Besides all this, other by-products of the analyses are a better understanding of air-shower physics, including for instance the effects of the geomagnetic field or the impact in the observed signals of atmospheric changes of the pressure or air  density. Also the detectors are used to study space weather effects, such as Forbush decreases which affect the background counts, or to study unusual lightning effects, such as the high altitude Elves observed with FD or some very extended signals in SD associated to thunderstorms.

\section{Results}

\begin{figure}[t]
\centerline{\epsfig{width=4in,angle=0,file=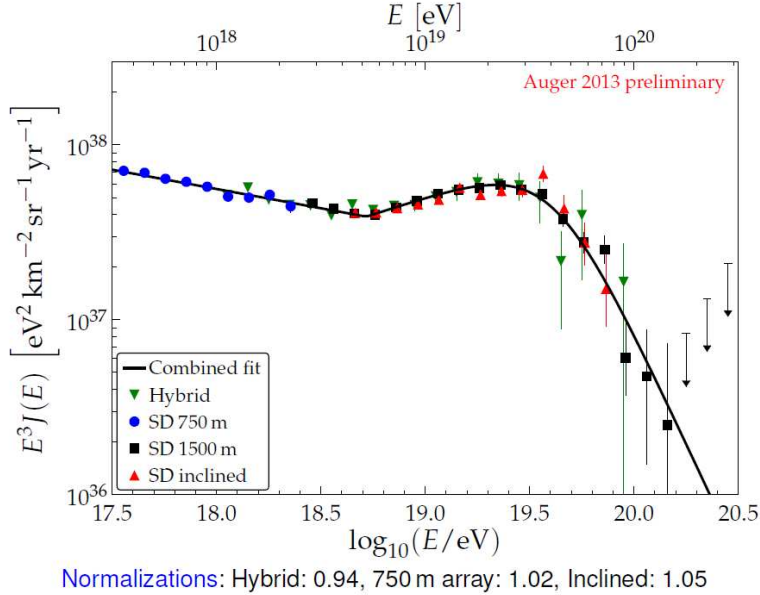}}
\vskip 1.0 truecm
\caption{CR spectrum obtained with the Infill, the hybrid data and the main array (both with vertical and inclined samples), slightly rescaled with the indicated factors, which are compatible with the systematic uncertainties, so as to improve their matching \cite{sc13}.} 
\label{fig1}
\end{figure}

The CR spectrum has been accurately determined at the Auger Observatory from $3\times 10^{17}$~eV up to beyond $10^{20}$~eV \cite{sc13}, showing two main characteristic features: the so-called ankle at $\sim 5$~EeV and a strong suppression above 40~EeV (see figure~1). The ankle represents the energy at which the spectral slope  $\alpha$ (where the differential flux is d$J/{\rm d}E\sim E^{-\alpha}$) changes from $\alpha\simeq 3.23$ to   $\alpha\simeq 2.63$. It is still not clear whether this change is related to a transition from  a Galactic component to a harder extragalactic one, or if instead this transition happens at somewhat lower energies and the ankle is explained for instance by the so-called dip model. In this last scenario the spectral hardening is the imprint left on  an extragalactic proton component by the attenuation due to  pair production processes in the interactions with the CMB background, for which the threshold would be at about the observed ankle energy.
The high-energy suppression is also clearly seen in figure~1, being the significance of the departure from the extrapolation of the power-law observed at lower energies of more than 20$\sigma$. The interpretation of this suppression is still not clear, because it could be related to the predicted GZK suppression  of a proton component attenuated by photo-pion processes off the CMB or to the attenuation of heavy nuclei, such as Fe, by photo-disintegration effects. It could also be affected by a limiting maximum acceleration at the sources themselves, which could not only modify the scenarios just mentioned but could also for instance allow for the so-called mixed scenarios, in which a maximum rigidity cutoff in the acceleration at the sources could lead to maximum energies for different nuclei scaling with the atomic number $Z$, with $E_{max}\simeq 5Z$~EeV.

\begin{figure}[t]
\centerline{\epsfig{width=5.7in,angle=0,file=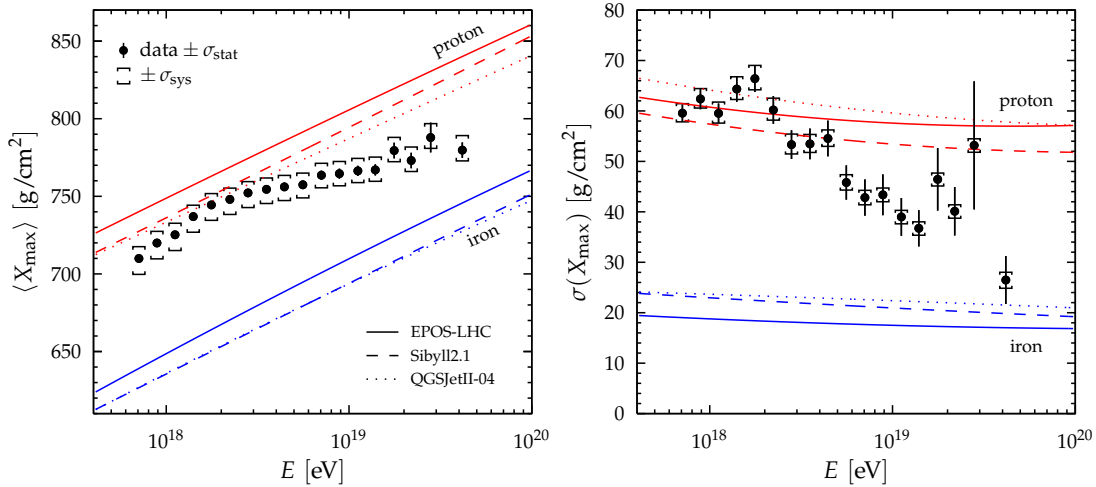}}
\vskip 1.0 truecm
\caption{$X_{max}$ and $\sigma(X_{max})$ as a function of energy \cite{xmax}.} 
\label{fig2}
\end{figure}

These maximum rigidity scenarios would also lead to a transition from a light to a heavy composition for increasing energies above $\sim 5$~EeV.
This kind of transition in the composition is indeed suggested by the measurement of the depth of shower maximum \cite{xmax}, as shown in figure~2.
Both the trend of $\langle X_{max}\rangle$, which for increasing energies increases more slowly than what would be expected from an unchanging composition, as well as the observed dispersions on its values, $\sigma(X_{max})$, which  for energies above the ankle become smaller than the values expected for pure protons, suggest that a transition towards a heavier composition takes place above the ankle\footnote{Note that a mixture of two different nuclei would lead to values of $\sigma(X_{max})$ larger than those expected for each individual nucleus, and hence the values observed above the ankle also suggest that there is not too much admixture between nuclei of very different masses.}.

Figure~3 shows the results of a likelihood fit to a scenario with four mass components ($p$, He, N and Fe), displaying the fraction of each element obtained in the different energy bins considered \cite{xmaxcomp}.  It is apparent that the proton component is quite important at EeV energies, which puts some tension to the models in which CRs  would be of Galactic origin up to the ankle. The fraction of protons then becomes significantly suppressed above $\sim 5$~EeV, while the fraction of He and N increase above that energy. The heaviest nuclear component, Fe, is seen to be very suppressed at all energies, even in the highest energy bin corresponding to $E>30$~EeV. For more details on composition results see  the talk dedicated to this subject at this meeting \cite{jarne}.

\begin{figure}[t]
\centerline{\epsfig{width=6in,angle=0,file=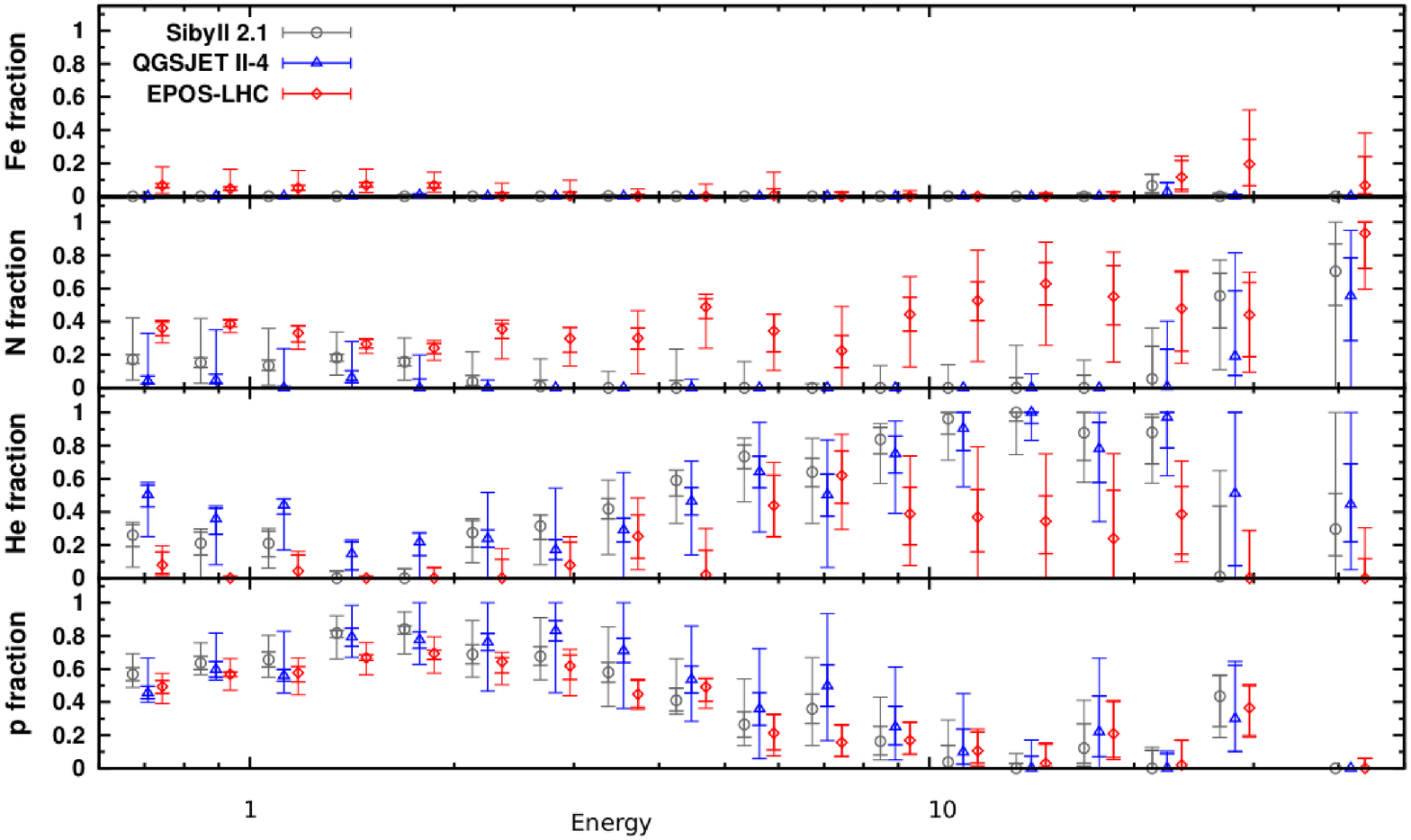}}
\vskip 1.0 truecm
\caption{Likelihood fit to a four component mass composition adopting different hadronic models \cite{xmaxcomp}.} 
\label{fig3}
\end{figure}

Searches for photons and neutrinos have been performed but with negative results, hence setting upper-bounds on their diffuse fluxes.  Photon induced air-showers
 are mostly electromagnetic, lacking the muons that tend to give shorter rise-times to the signals in the WCDs. Being also more penetrating than hadronic showers of the same energy, they tend to have a more curved shower front. These characteristics are used to identify the photon candidates and allowed us to set the bounds on the fraction of the fluxes contributed by photons \cite{photons} which are shown in Fig.~4a. These bounds already exclude many exotic candidates for UHECR production in top-down models that were proposed in the past, such as those involving decays of super-heavy dark matter, decays of topological defects, etc., whose predictions would exceed the bounds obtained. With the data to be collected in the next few years, and also exploiting improvements in the triggers recently implemented, it will become possible to test some of the scenarios of photon production in  association with the GZK effect if protons from far away sources  do contribute to the fluxes at the highest energies (colored bands in the plot, while the solid lines illustrate the sensitivities that could be achieved as early as 2015).

\begin{figure}[t]
\centerline{\epsfig{width=3.1in,angle=0,file=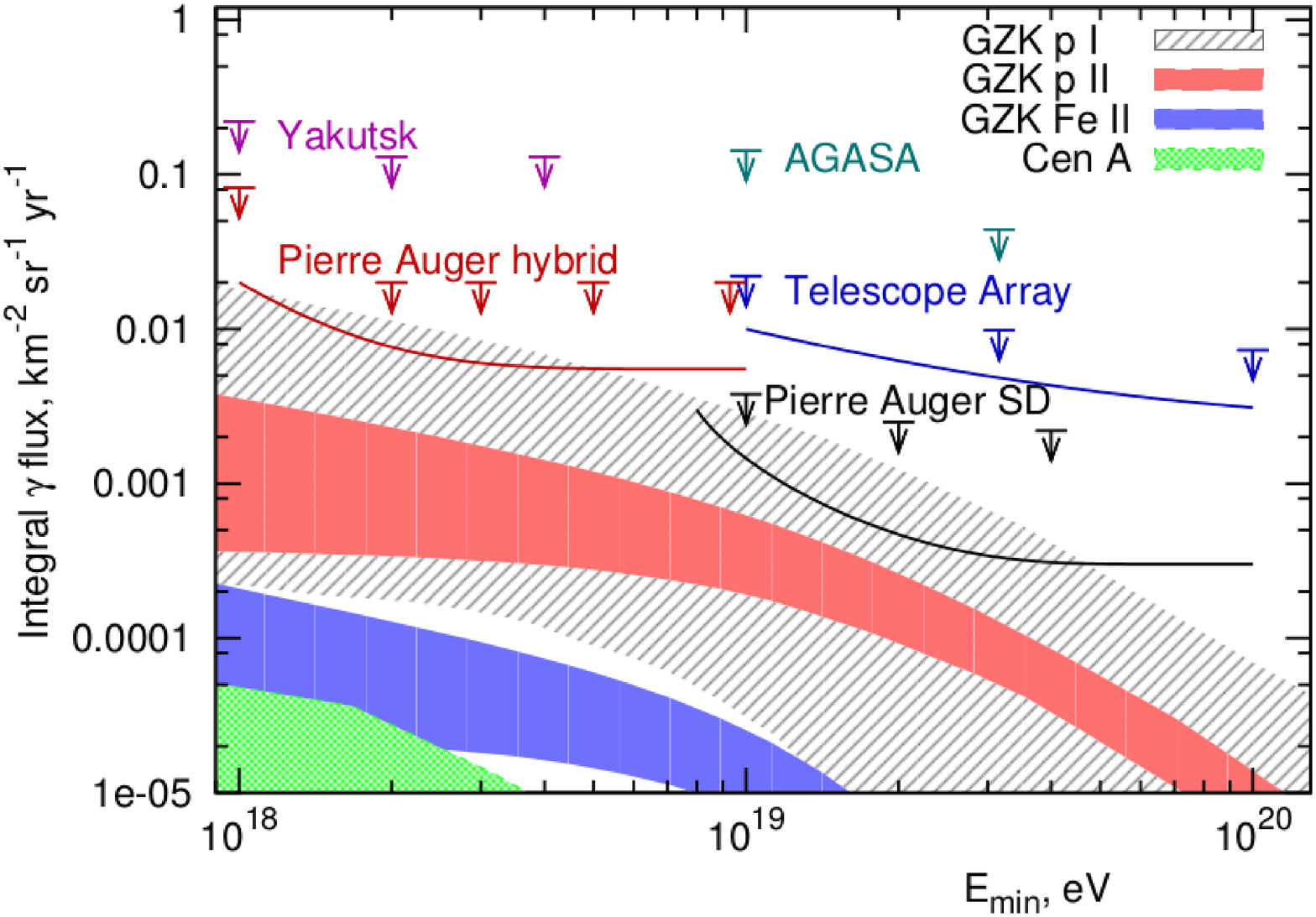}}
\centerline{\epsfig{width=2.8in,angle=0,file=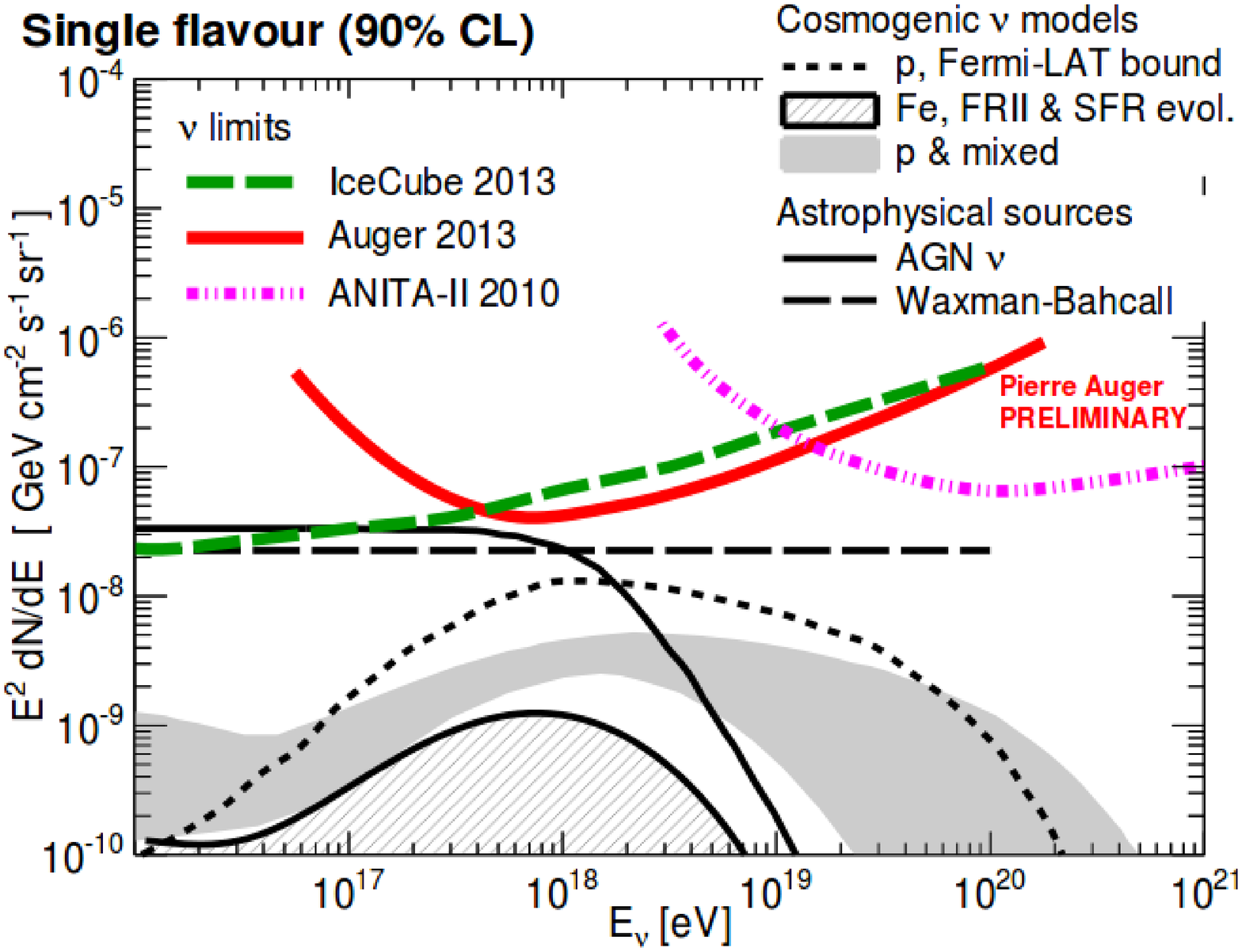} }
\vskip 1.0 truecm
\caption{Upper bounds on diffuse fluxes of photons \cite{photons} (top-panel) and neutrinos \cite{neutrinos} (bottom-panel).} 
\label{fig4}
\end{figure}

Neutrinos, being so weakly coupled, can interact in the atmosphere essentially at any depth, contrary to hadronic showers that are initiated in the first $\sim 100$~g/cm$^2$ of the upper atmosphere.  Hence, looking at very inclined showers close to the horizon the large background of hadronic showers can be clearly separated from the neutrino induced showers initiated close to the detector. These last will have a young electromagnetic component  while the corresponding electromagnetic component of the hadronic showers  produced very far away  would have been almost completely absorbed by the atmosphere. Actually, the most efficient way to identify neutrinos is  through the $\nu_\tau$ component (one expects that due to oscillations all flavors will be equally represented in the fluxees from far away sources). Tau neutrinos grazing the surface of the Earth, i.e. arriving from slightly below the horizon, may produce $\tau$ leptons in charged current interactions in the rock, and these can travel several tens of km before decaying. If they decay hadronically after coming out from the ground and do so near the area of the Auger Observatory  they will produce showers that can be detected. The lack of observation of neutrino induced events allowed us to set the bounds on the diffuse fluxes presented in Fig.~4b \cite{neutrinos}, which are still slightly above, but not far, from the optimistic expectations of cosmogenic neutrinos produced by a protonic component of the UHECR interacting with the CMB. It is also important to keep in mind that if the CR composition indeed becomes heavier, such as in the mixed scenarios with a
rigidity dependent cutoff with $E_{max}\simeq 5 Z$~EeV, the fractions of nucleons with  $E_{nucl}> 10$~EeV  will be very strongly suppressed and hence the neutrinos they can produce in photopion processes, which typically have $E_\nu\simeq E_{nucl}/20$, will give negligible fluxes above 0.5~EeV. This is where the Auger Observatory has greatest sensitivity.  Hence, to  observe cosmogenic neutrinos, as well as GZK photons,  it is crucial to know whether the fraction of protons above 10~EeV is non-negligible. 

Regarding the study of the distribution of arrival directions, two main types of searches have been performed, that of anisotropies on large angular scales in the form of a dipole or a quadrupole, and  of anisotropies on small or intermediate angular scales, typically on angular windows with radius from $1^\circ$ up to 30$^\circ$.

\begin{figure}[t]
\centerline{\epsfig{width=3in,angle=0,file=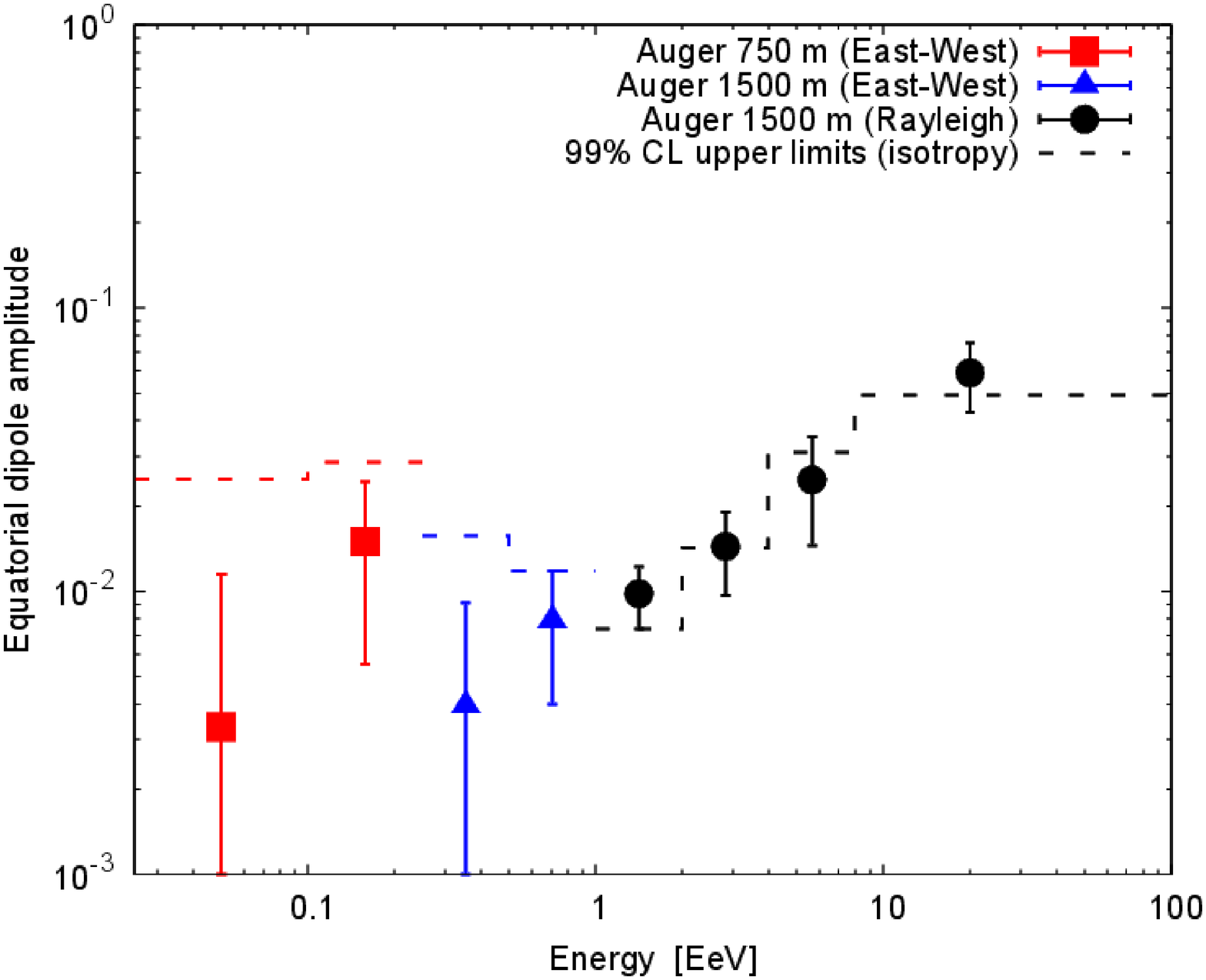}}
\centerline{\epsfig{width=3in,angle=0,file=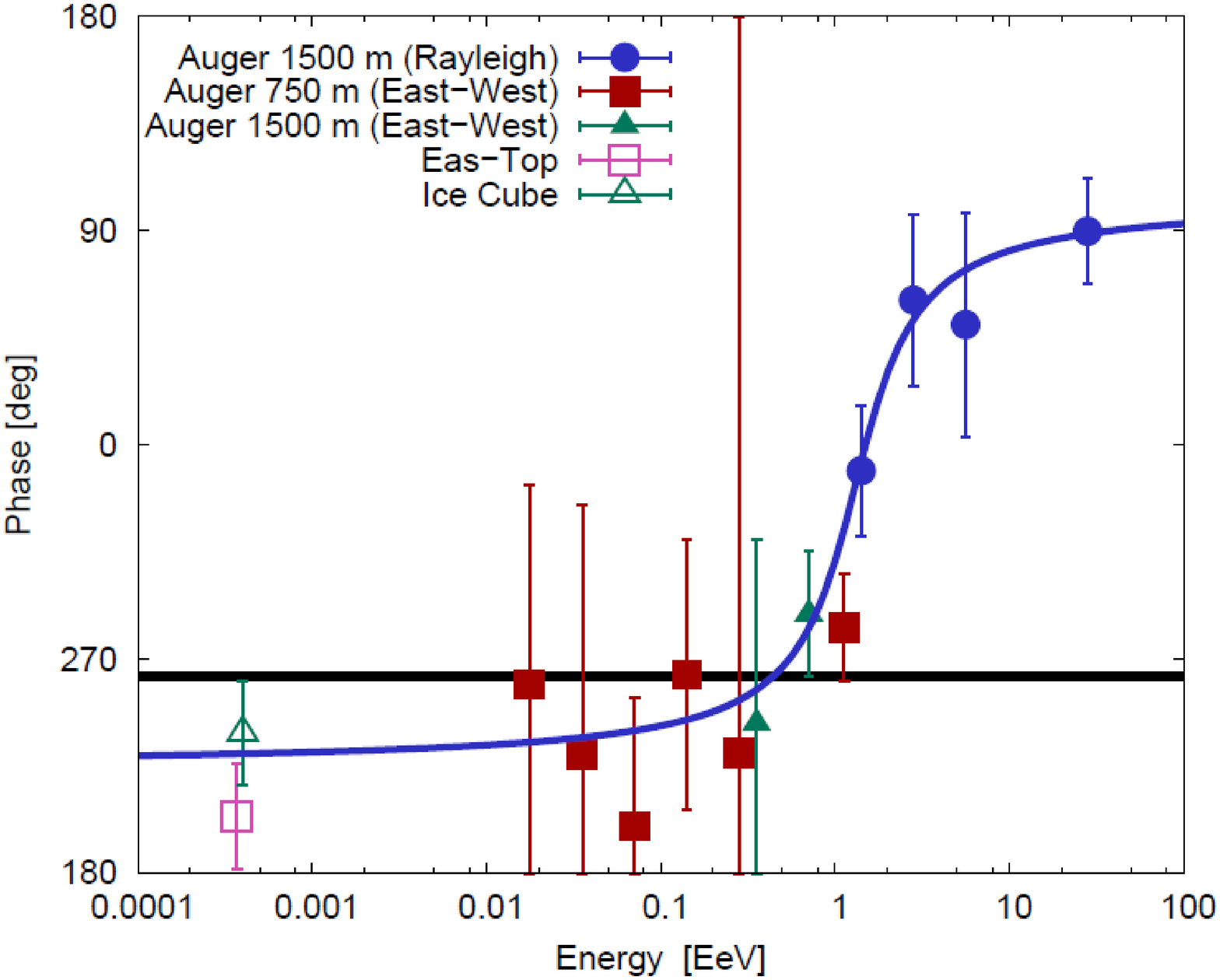}}
\vskip 1.0 truecm
\caption{Rayleigh amplitude (top) and phase (bottom) as a function of energy. A smooth fit to the phases is just included to guide the eye \cite{rayleigh13}.} 
\label{fig5}
\end{figure}

The simplest large-scale analysis consists in the study of the distribution in right ascension to obtain the Rayleigh amplitude and phase, which are related to the equatorial component of the dipole. On the other hand, the full distribution in right ascension and declination can be obtained through a multipolar analysis or by combining Rayleigh analyses in the right ascension and azimuthal distributions, this last being sensitive to the component of the dipole along the axis of rotation of the Earth. One needs however to control several sources of systematics to achieve the required precision, such as the non-uniform exposure in time (accounting for the actual active size of the array at any given moment), correct the energy assignment for the effects of the atmospheric variations and, in order to obtain the whole dipole in the sky, also to account for a slight tilt of the array (of 0.2$^\circ$ on average) and correct for the effects of the geomagnetic field on the muonic part of the air shower, which could otherwise induce a spurious 
N-S dipolar component. In addition, for energies below full efficiency, where trigger effects could also lead to spurious modulations, it is possible to obtain the Rayleigh coefficients by the so-called East-West method. This method  recovers the modulation in right ascension essentially from the value of its derivative, estimated from the difference between the counts towards the East and the West at any given time. These are equally affected by the systematic effects previously mentioned and hence allow us to obtain a cleaner measurement although with somewhat reduced sensitivity.
The results presented at the ICRC 2013 are shown in figure~5 \cite{rayleigh13}, which includes events with zenith angles up to 60$^\circ$ measured before the end of 2012. One can see that 3 out of the 4 bins above 1~EeV have an equatorial component of the dipole (inferred from the Rayleigh amplitude and the average declination of the events) larger than the expectations for 99\% of the isotropic simulations, indicated with a dashed line. 
Another characteristic feature of the results is that the Rayleigh phases in all bins below 1~EeV point approximately towards $RA\simeq 270^\circ$, which happens to be quite close to the right ascension of the Galactic Center, while the phase changes above 1~EeV making a transition towards $RA\simeq 90^\circ$ above 2~EeV. This may be indicative of a transition from a Galactic to an extragalactic CR population in this energy range. A puzzling feature is that the composition at 1~EeV already seems to have a significant light (proton-like) component but, if the  protons at these energies were Galactic, one would  expect, for typical models of the Galactic magnetic fields, that they would lead to quite large anisotropies as they escape from the Galaxy. In particular, both the dipolar and the quadrupolar components would be larger than the bounds obtained if a significant component of Galactic protons is present at 1~EeV, and hence those protons are most likely already extragalactic.

\begin{figure}[t]
\centerline{\epsfig{width=5in,angle=0,file=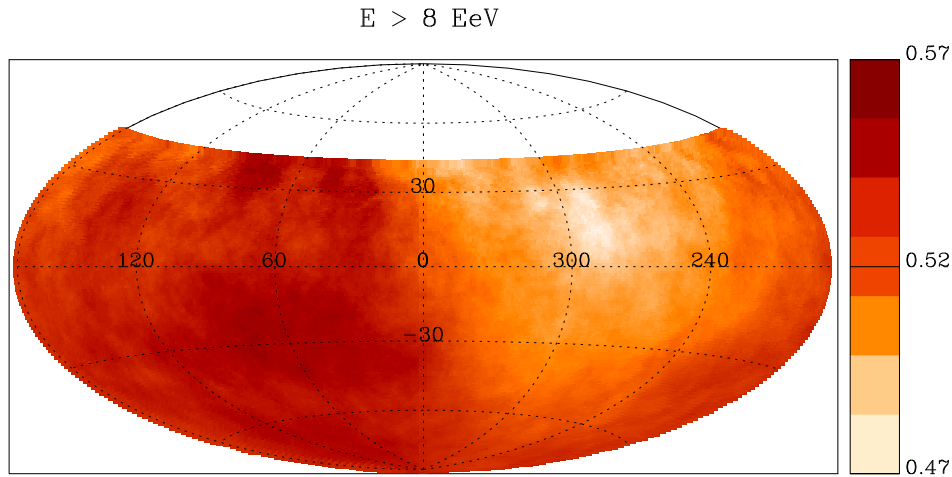}}
\vskip 1.0 truecm
\caption{Map of the CR flux for $E>8$ EeV averaged on 45$^\circ$ windows (in km$^{-2}$yr$^{-1}$sr$^{-1}$) \cite{ls14}.} 
\label{fig6}
\end{figure}

A recent analysis of large scale anisotropies \cite{ls14}, including data up to the end of 2013 and incorporating the `inclined' events with zenith angles between 60$^\circ$ and 80$^\circ$, which are reconstructed using a different method than the one used for `vertical' sample with $\theta<60^\circ$, obtained the dipole and quadrupole components in the energy bins [4,8]~EeV and $E>8$~EeV, for which both samples are above full trigger efficiency. Figure~6 shows a map of the CR flux, smoothed on circular windows of 45$^\circ$ radius, for the events with $E>8$~EeV, for which a dipolar modulation in the distribution is quite apparent. In this energy bin the significance for the right ascension amplitude obtained, $r=(4.4\pm 1.0)\times 10^{-2}$, exceeds 4$\sigma$.
This dipolar  modulation, likely of extragalactic origin, may be due to the non-isotropic distribution of the sources in our local neighborhood (for instance, the non-isotropic distribution of galaxies within $\sim 90$~Mpc is known to be responsible for our peculiar motion with respect to the CMB, and a similarly anisotropic distribution of CR sources could give rise to a significant large scale anisotropy).
The dipolar anisotropies may also be affected by the diffusion of the CRs in the intergalactic turbulent magnetic fields if these are sizeable.

\begin{figure}[t]
\centerline{\epsfig{width=2.2in,angle=0,file=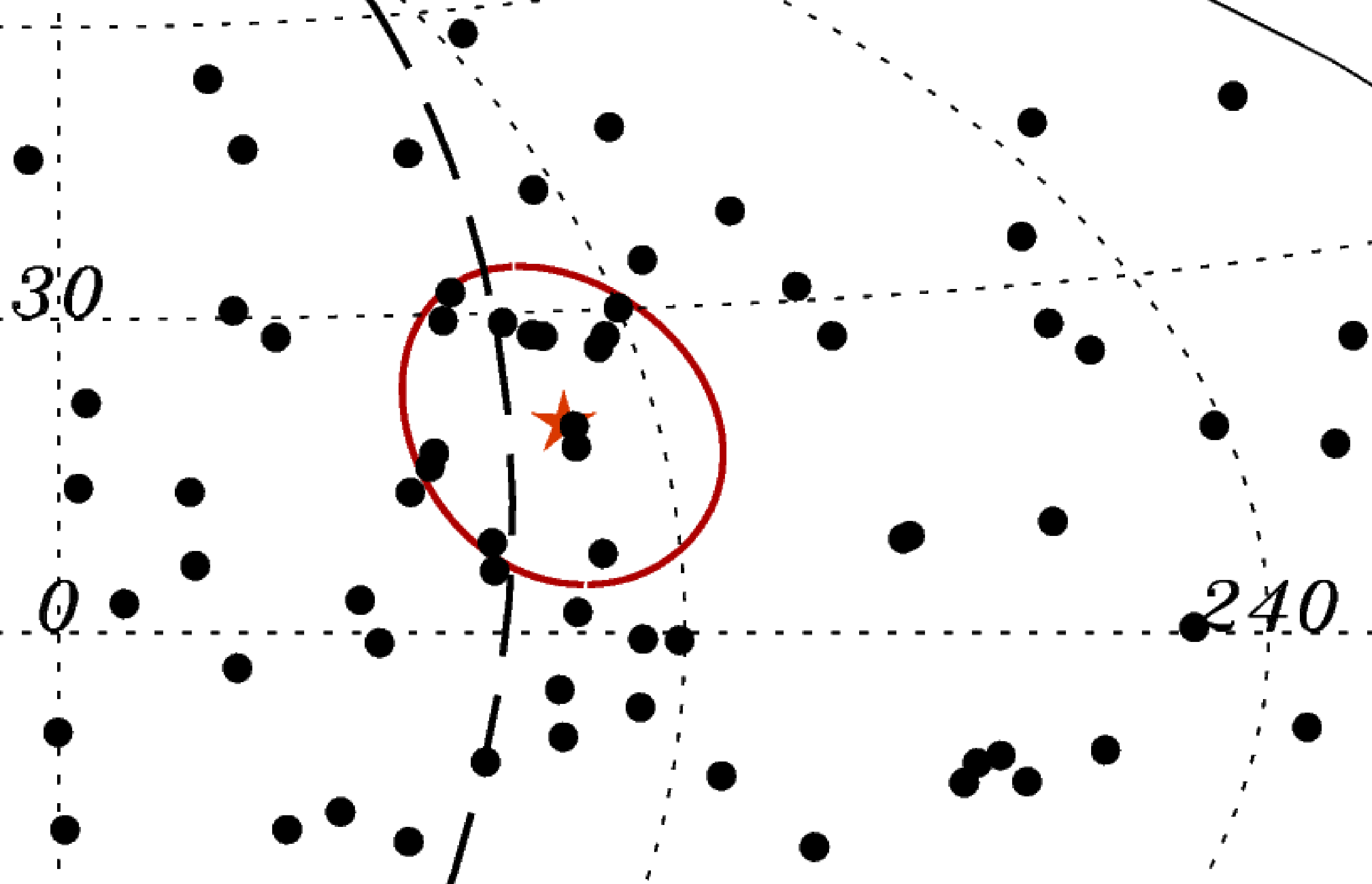}}
\vskip 1.0 truecm
\centerline{\epsfig{width=3.1in,angle=0,file=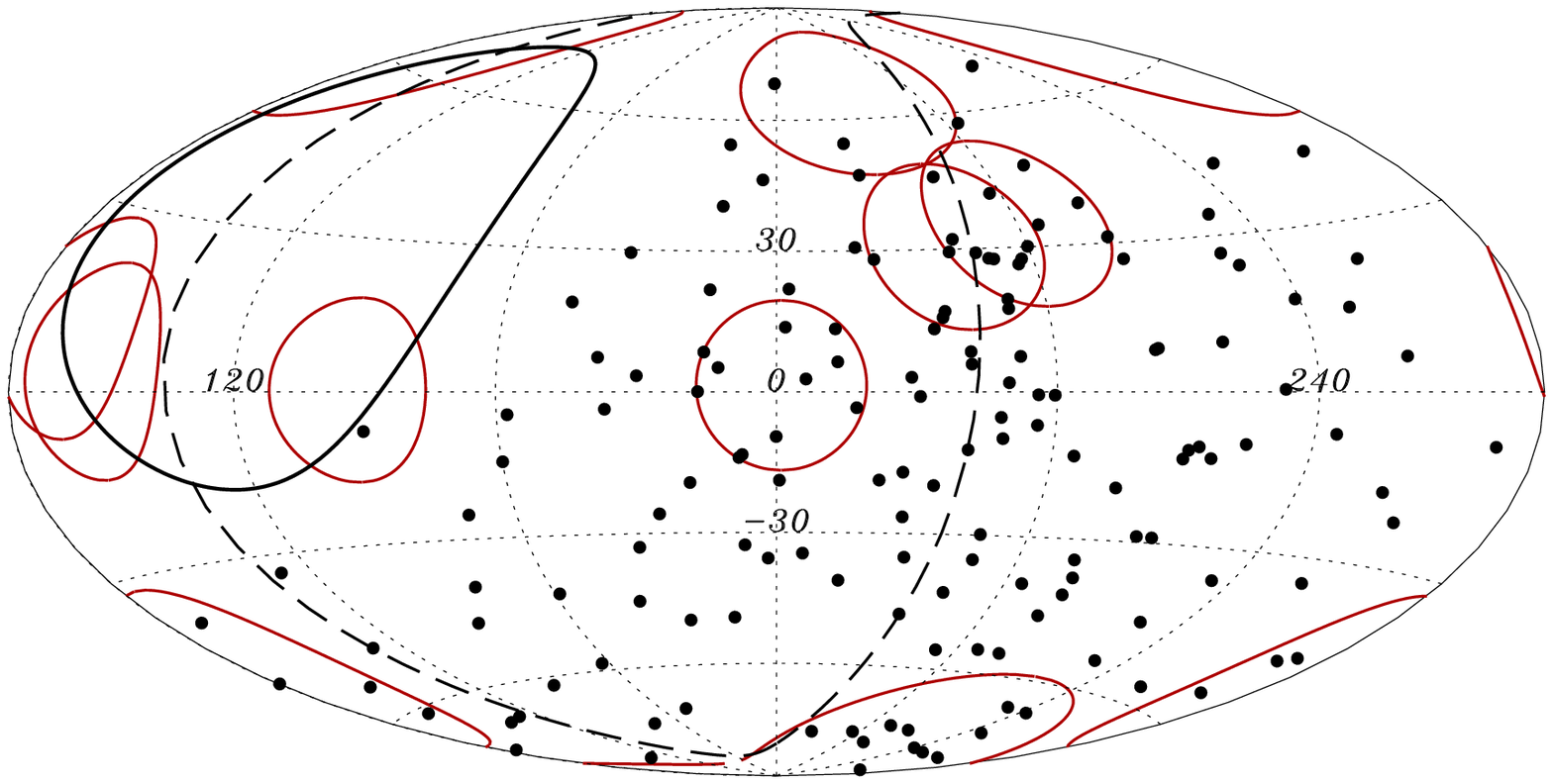}}
\vskip 1.0 truecm
\caption{Top: Map of events with $E>58$ EeV in the region around Cen~A (star with a circle of 15$^\circ$ radius). Bottom: sky map with 18$^\circ$ radius circles around the 10 brightest SWIFT AGNs within 130~Mpc \cite{aniso14}.} 
\label{fig7}
\end{figure}

The search for anisotropies at smaller angular scales and for the highest energies, $E>40$~EeV, for which the magnetic deflections may be of only a few degrees for protons and for which only relatively nearby sources (closer than $\sim 200$~Mpc) may contribute due to the GZK effect, has been recently updated using the whole vertical and inclined samples up to end of March 2014, corresponding to a total exposure of about 66,000 km$^2$ yr sr \cite{aniso14}. This sample contains 602 events above 40~EeV and 231 with $E>52$~EeV. Several tests were performed, both looking at intrinsic anisotropies on the arrival directions, such as localized excesses on circular windows in any direction of the sky  or looking at the autocorrelation among the events, and also looking at the cross-correlation with different catalogs, with the Galactic and Super-Galactic planes or  with the direction towards Centaurus~A (Cen~A), which is the closest Active Galactic Nucleus (AGN). In both cases a scan on the energy threshold from 40~EeV up to 80~EeV was performed, as well as a scan on the angular size considered from $1^\circ$ to $30^\circ$.

The catalogs we considered were the 2MRS IR galaxy catalog (which could be tracing for instance the location of GRBs or of fast spinning newborn pulsars), the AGN in the SWIFT BAT 70 month X-ray  catalog (dominated by Seyfert active nuclei hosted mostly by spiral galaxies) and a catalog of radiogalaxies with jets (dominated by Fanaroff-Riley AGNs hosted mostly by elliptical galaxies). In the case of the AGN catalogs, besides considering the whole flux limited samples we also considered subsamples defined by a threshold in the intrinsic luminosity, motivated by the fact that more luminous AGNs have in general stronger associated  magnetic fields and hence may also be able to accelerate CRs up to larger maximum energies.
For the catalogs we also scanned on the maximum distance to the objects considered, from 10~Mpc up to 200~Mpc. 

The overall distribution of arrival directions turns out to be  remarkably 
isotropic, with for instance 70\% of isotropic simulations having an autocorrelation of events with a more anisotropic signature than the actual data under a similar scan. The events indeed fill most of the observed regions of the sky.  This could be interpreted as originating from the fact that the UHECR tend to be heavy at these energies and hence the deflections in the Galactic and extragalactic magnetic fields could be significant, and/or the number of contributing sources could be quite large, so that no individual source contributes a large fraction of the observed flux. 

From the remaining analyses, the largest observed departures from isotropy are those obtained  when considering angular windows of 15$^\circ$ around the direction towards Cen~A for events with $E>58$~EeV, for which 14 events are observed while 4.5 were expected on average for isotropic distributions. The other case is when considering X-ray AGNs from SWIFT brighter than $10^{44}$~erg/s and closer than 130~Mpc. In this case 62 pairs between them and the events with $E>58$~EeV are formed within $18^\circ$ while only 32.8 were expected on average to arise by chance coincidences in isotropic distributions.  Both results have penalized probabilities (accounting for the different scans performed) close to 1\%. The map of events with $E>58$~EeV around the direction towards Cen~A is shown in the top-panel of figure~7, while the whole sky map of the events with the location of the 10 brightest AGNs within 130~Mpc, indicated with circles of 18$^\circ$ radius, is shown in the bottom-panel. It is important to monitor these excesses to see if they get confirmed with an independent data set.

Regarding the future, the Auger Observatory is expected to continue operating for another decade and moreover an upgrade is foreseen in which the electronics of the WCDs will be improved and each WCD will be supplemented with a 4~m$^2$ scintillator detector on top of it. This should allow the separate reconstruction of the electromagnetic and muonic components of each shower and hence provide an improved determination of the CR composition using the SD. This should also allow to reach a better understanding about the modelling of the hadronic interactions at the highest energies.  In addition, in the Infill region scintillators buried underground are being deployed (AMIGA detectors) in order to measure directly the muonic component of the showers.

All this should allow to  gather a larger quantity of data with an improved quality, so that one could expect that further advances in the field will be achieved in the near future.

\section*{Acknowledgments}
 Work supported by CONICET and ANPCyT, Argentina.

\end{document}